\documentclass[12pt]{iopart}

\usepackage{iopams,graphicx} 

\begin{document}

\title{ Reduction of magnetic perturbation for SR-POEM}

\author{B R Patla }

\address{Smithsonian Astrophysical Observatory,\\
 Harvard-Smithsonian Center for Astrophysics,\\
60 Garden St, Cambridge, MA 02138, USA 
}
\ead{bijunath.patla@nist.gov}

\begin{abstract}
SR-POEM is a Galilean test of the weak equivalence principle(WEP) that aims to measure the fractional acceleration difference $\eta$  with a mission uncertainty $\sigma(\eta)=1\times 10^{-17}$for a pair of test substances. It is to be conducted during the low-drag free fall portion of a sounding rocket flight. 
The interaction of the  magnetic field gradient with tiny remanent magnetic moment of the test masses (TMs) will produce a spurious acceleration that is not sufficiently reduced by a single Mu-metal shield. In this paper, we study configurations with two and three shields. 
Approximate analytic formulae are used to study the shielding factor as a function of geometry.
We use finite element analysis (FEA) to determine the magnetic field and gradient for certain cases that fit the mission requirements. FEA results are compared with analytic expressions wherever appropriate.
Several configurations reduce both axial and transverse magnetic field by at least the required factor of $4\times 10^5$. 

\end{abstract}


\maketitle

\section{Introduction}
SR-POEM is a Galilean test of the WEP which is to be conducted during the low-drag free fall portion of a sounding rocket flight \cite{rdr1, rdr2, rdr3}. We compare the rate of fall of two test masses (TMs) that contain two different test substances and that are monitored by a set of precision laser gauges \cite{jdp, thapa}. Our recent investigations into magnetic disturbances have led us to conclude that $\eta$, the fractional acceleration difference, can be measured with an uncertainty $\sigma(\eta)\le 1\times 10^{-17}$, a 10,000-fold advance in the state of the art. Payload inversions are central to the reduction of systematic error.

Substances possessing an intrinsic(remanent) or induced magnetic moment interact  with the magnetic field gradient. This interaction results in a force on the SR-POEM TMs (see \cite{rdr4} for a detailed discussion of SR-POEM test mass design) is by far the largest contributor to the error budget. Below, we consider the reduction of moment; but we conclude that substantial reduction of gradient is also necessary. The sources of most SR-POEM errors are fixed relative to the payload, so they cancel with inversion. The Earth's magnetic field is fixed relative to the Earth. We show that relevant components (those along the axis of measurement) of the magnetic force do not cancel with inversion. The main part of this paper discusses shield configurations and establishes few satisfactory ones. 

Local sources of magnetic field and gradient at the TMs can be reduced by using materials with low magnetization. For unavoidable magnetic fields from payload components, e.g., those from an ion (vacuum) pump, magnetic flux return paths can be provided. Shielding is still needed to further reduce these fields. The gradient of the Earth's magnetic field above 1500 km and away from magnetic materials is quite small, $\sim
 10^{-11}$\,T/m. Ironically, the gradient of the Earth's (attenuated) field inside the necessary shield is the largest field gradient for SR-POEM. 

\section{Magnetization of the TMs \label{sec1}}

Although the TMs are made of nominally non-magnetic materials, the presence of ferromagnetic impurities can result in a permanent moment. Below, we discuss the permanent magnetic moment of Al, one of the test substances for SR-POEM. The moment of all TMs will need to be tested.
In the following paragraphs we summarize the results of some studies that are relevant to SR-POEM test masses. 

Su \etal(1994) estimated magnetic moments of Al and Be test masses using a torsion pendulum, by measuring the twist angle of the pendulum tray with test masses in an ambient magnetic field. Based on the data in their paper, we obtain a moment $m=5 \times 10^{-10}$\;A\;m$^{2}$ for their test mass (table\,\ref{tab1}). They also report a value of  $2.8\times10^{-8}$\;A\;m$^{2}$ for the permanent magnetic moment of a 50\;g Al tray. 
Apparently the tray had ferromagnetic impurities\,\cite{gundlach-p}. 
We will ignore the Al tray for the remainder of this paper because by  testing and selecting materials, we expect to avoid such anomalies.

Mester and Lockhart (1996) measured a magnetic moment $\sim 4.0\times 10^{-10}$\;A\;m$^{2}$ on Al samples of 6\;mm diameter and 6\;mm length.
These samples may have contained a significant iron content picked up during  machining, from the tools or from contaminated cutting 
fluids\,\cite{mester-p}. 

More recently, the LISA technology package proof mass (73\%Au--27\%Pt, $\chi=-2.5\times 10^{-5}$) weighing a kilogram, with sides of length 0.04 m, had a measured magnetic moment of $2.0\times 10^{-8}$\;A\;m$^{2}$\,\cite{hueller-p,hueller}. The sensitivity of the measuring instrument was comparable to the measured values and the moment was observed to vary after the test mass was touched (creating a thermal gradient of $\sim 30~K$ with a time constant of $\sim$ 100s). Thermal emfs were identified as a likely cause. 

In contrast to this, SR-POEM TMs will be subject to temperature changes and differences across the TM housing of only $\sim 1 \mu{}K$. SR-POEM TMs---cut from large pieces---are of a more uniform composition(thermal conductivity depends ordinarily on temperature, pressure and composition)than the LISA TMs, which were cast individually. Although thermally-induced moments seem unimportant in SR-POEM TM, we plan to test all TMs and measure any remanent 
magnetization.

We considered the magnetic moment measurements (of Al and similar metals or alloys) reported by these authors and scaled the results to the SR-POEM test mass volume of $4.0\times 10^{-4}$\,m$^3$ (table\,\ref{tab1}. At the expense of being conservative, we decided to go with a geometrical scaling (for SR-POEM TM that is volume) that yields higher value for the magnetic moment.
For the remainder of this paper, we will assume the moment of the TMs to be $5.0\times 10^{-8}$\,A\,m$^{2}$, slightly larger than the volume-scaled dipole moment of the Al test masses based on the data of Su\,\etal\,(2004). The moment of SR-POEM TMs will need to be tested to at least this sensitivity.

\Table{\label{tab1} Comparison of measured magnetic moments of test mass samples by various authors available in published literature. These measurements have been scaled to match with SR-POEM test masses.}
\br
&&&\centre{2}{Magnetic moment (A\,m$^{2}$)}\\
\ns
&&&\crule{2}\\
Authors& Volume\,(m$^3$) & Material & Measured  & Scaled$^{\rm a}$\\
\mr
Su et al.(1994)                     &$\quad 5.4\times 10^{-6}$  & Be-Al$^{\rm b}$    & $5.1\times 10^{-10}$ &$3.7\times 10^{-8}$\\
Su et al.(1994)                     &$\quad 1.9\times 10^{-5}$ & Al tray               & $2.8\times 10^{-8}$ &$6.1\times 10^{-7}$\\
Mester \& Lockhart (1996)           &$\quad 1.7\times 10^{-7}$ & Al                 & $4.0\times 10^{-10}$ &$9.4\times 10^{-7}$\\
Hueller (2010)                      &$\quad 6.4\times 10^{-5}$ & Pt/Au$^{\rm c}$             & $2.0\times 10^{-8}$  &$1.3\times 10^{-7}$\\

\br
\end{tabular}
\item[] $^{\rm a}$ Magnetic moment scaled to the volume of an Al test mass of SR-POEM of mass $\sim 1$\,kg. Although we have assumed that the magnetic moment scales with volume, it is quite possible that the moment is due to impurities on the surface and that the correct scaling is with area. In that case, the scaled moments will have a lesser value, proportional to the square of the ratio of the dimensions instead of the cube.
\item[] $^{\rm b}$ A comparison of the moments of separate Al and Be test masses, not a measurement of the moment of an alloy.
\item[] $^{\rm c}$ LISA test mass itself is an alloy (Pt-Au, 27\%--73\%) and we have approximated the density of the test mass by that of Au.
\end{indented}
\end{table}

\section{Force acting on the TMs \label{sec-force}}
A magnetized TM may be approximated as a magnetic dipole. In a magnetic field $\vec{B}$ (inside the shield), the force acting on the TM is
\begin{equation}
\vec{F}=\vec{\nabla}(\vec{m}\cdot\vec{B}),
\label{force}
\end{equation}
where $\vec{m}$ is the TM dipole moment, which has  permanent and induced components, $\vec{m}=\vec{m}_p+\vec{m}_{in}$, where $\vec{m}_p$ and $\vec{m}_{in}$ are the permanent and induced moments. If we assume that the TMs are magnetically isotropic, the induced moment should be parallel to the applied field. We have not included higher order moments in our analysis, although they could be important for spatial separations that are small compared to the dimension of the TM.  We note that the LISA test masses---at any instant could be approximated with a single dipole with a rotating direction and changing amplitude\cite{hueller-p}.

$\vec{m}_{in}=\chi V \vec{B}/\mu_0$, where $\chi$ and $V$ are the susceptibility and volume of the TM and $\vec{B}\sim \vec{B}_E/S_F$, where $\vec{B}_E\approx 5.0\times 10^{-5}$~T is an assumed value for the appropriate component of Earth's magnetic field and $S_F\approx 1000$  is an assumed shielding factor.
Noting that the susceptibility of Al is $\chi=2.2\times 10^{-5}$, we obtain a value for the induced moment of $m_{in}\approx 3.2\times 10^{-10}$\,A\,m$^2$,
which is approximately two orders of magnitude below the expected value for the permanent moment. Moreover, the force due to the induced moment cancels after inversion as shown in figure\,\ref{fig1}---a schematic diagram representing a single TM.
Therefore, except while discussing inversion below, we will ignore the induced moment and consider only the contribution of the permanent moment to the magnetic force. With no {\it a priori} information about the direction of the permanent moment, we set equal requirements on its axial and transverse components. 

We adopt Cartesian coordinates in which the  $z$ direction  coincides with the payload's symmetry axis, along which the WEP measurement is performed, see figure\,\ref{fig1}. Expressing equation \eref{force} in component notation we have $F_i=\partial_i(m_j B^j)$, where repeated indices are summed from 1 to 3. For the $z$ component, 
\begin{equation}
F_z=2B_j\frac{V\chi}{\mu_0}\partial_z B^j+(m_p)_j\, \partial_z B^j.
\label{force1}
\end{equation}
The field external to the shield is the Earth's field; we neglect all other sources  as they are small compared to Earth's magnetism. The payload is nadir-pointing during WEP measurements. We decompose the Earth's field into its vertical component, which is axial, and its horizontal component, which is transverse with reference to the payload. Normally, shielding axial fields is more difficult, and gradients are higher than in the transverse case (table\,\ref{tab4}). 

To set the scene, we start by using a rough estimate for the magnetic field gradient $\partial{B_i}/\partial{z} \sim B_E/(S_F r_0)$ for computing the force on a TM. Here $B_E$ is the appropriate component of Earth's field, axial or transverse, $S_F$ is the appropriate shielding factor, and $r_0$ is a dimension characteristic of the shield which we take here to be the radius $\sim 0.1$~m.
The Earth's field at 1500 km above NASA's Wallops Island Flight Facility has a vertical component (axial, for SR-POEM) of about $2.4\times 10^{-5}$~T, and a horizontal (transverse) component of about $1.1\times 10^{-5}$~T. 
We concentrate on the larger value of the field and after imposing the requirement that the magnetic acceleration be 1/3 of the mission error, the force on the TMA must satisfy the condition
\begin{equation}
F \sim {(m_p)_i} \frac{\partial{B_i}}{\partial{z}} \le 1.0 \times 10^{-17}~\rm{N},
\label{force2}
\end{equation}
where $\partial{B_i}/\partial{z}$ is the gradient due to the transverse component of external field. With $(m_p)_i \le 5\times 10^{-8}$~A m$^2$, we require a shielding factor $S_F \ge 7.0 \times 10^{5}$. It is important to keep in mind that that the actual values of the maximum gradients depend on the shield geometry and configuration when multiple shields are used.

\subsection{Inversion and systematic error}
\begin{figure}[htb]
\begin{center}
\includegraphics[width=0.75\textwidth]{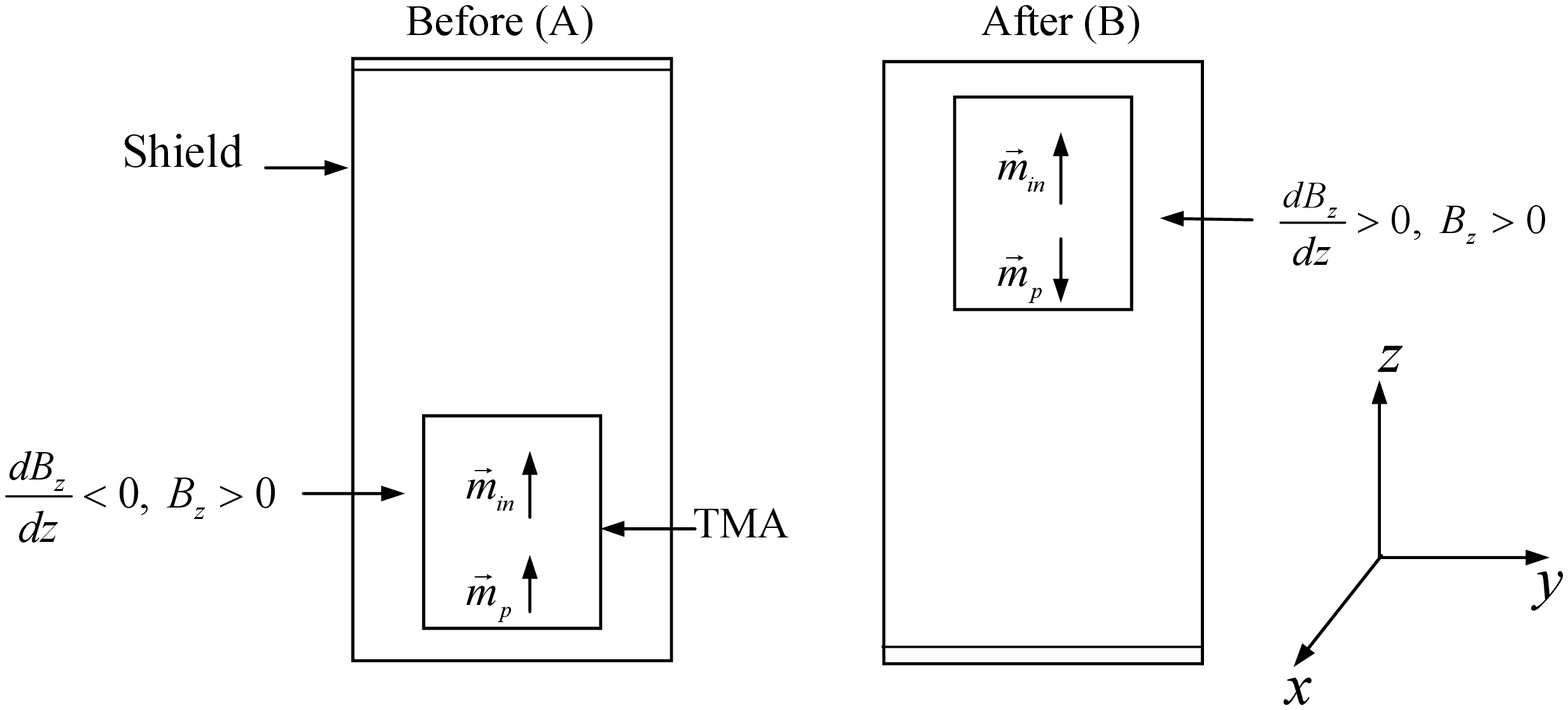}
\caption{Schematic of a TM before (A) and after (B) inversion, as observed from an inertial reference frame. The field inside the shield is due to the external field, so remains in the same direction. The TM's induced moment is along the direction of the external field whereas its permanent moment reverses with inversion. The magnitude of the field is a minimum at the center of a symmetrical shield, so the field gradient at the location of the TM reverses sign.
}
\label{fig1}
\end{center}
\end{figure}

In SR-POEM the payload is inertially pointed during a drop and is inverted between each successive pair of drops. When we add the differential accelerations of consecutive drops, calculated in an inertial reference frame, systematic error due to most sources cancel. Some components of the magnetic acceleration add instead of canceling because both the magnetic moment and the field gradient reverse sign on inversion. A schematic of a single TM as seen by an inertial observer before and after inversion is shown in figure\,\ref{fig1}. 

In figure\,\ref{fig1}, if the TM is placed symmetrical with respect to the shield, the $z$ component of the gradient is zero at the mid-plane. But the end caps are likely to be different and the field will not be uniform anymore at the location of the TM. Other symmetry-breaking effects are a variation of shield thickness or permeability with height, or positioning the TMs away from the center of the shield which may be necessary.
Thus, the gradient at the TMs is unlikely to be zero. 
If the TMs are off-center as shown in figure\,\ref{fig1}, we expect that the gradient approximately reverses sign upon inversion. Due to other various factors that lead to  asymmetries the reversal of the field gradient is not always guaranteed.
 Suppose the direction of the permanent moment $m_p$ is aligned with the direction of the external field before inversion. Then the force  acting on the TM in payload orientation A, in an inertial coordinate system, is
 \begin{equation}
 F_A=m_p\frac{dB}{dz}\biggm|_{A}+\,2 m_{in} \frac{dB}{dz}\biggm|_{A},
 \label{forceA}
 \end{equation}
 and in  orientation B it is
\begin{equation}
 F_B=(-m_p)\frac{dB}{dz}\biggm|_{B}+\,2 m_{in} \frac{dB}{dz}\biggm|_{B}.
 \label{forceB}
 \end{equation}
 Note that,
\begin{equation}
 \frac{dB}{dz}\biggm|_{A}=-\frac{dB_z}{dz}\biggm|_{0} +\delta h \quad {\rm and} \quad  \frac{dB}{dz}\biggm|_{B}=\frac{dB_z}{dz}\biggm|_{0} +\delta h,
 \label{marker}
 \end{equation}
 where $dB_{z}/dz|_{0}$ is the component of the gradient that changes sign with inversion and $\delta{}h$ is the component that does not. The Earth's field applied to a symmetrical shield with TMs off-center gives a gradient that changes sign with inversion. Using equation \eref{marker} in equations\,\eref{forceA} and \eref{forceB} yields
 \begin{equation}
 F_A=m_p\left( -\frac{dB_z}{dz}\biggm|_{0} +\delta h \right)+\,2 m_{in {\rm A}} \left( -\frac{dB_z}{dz}\biggm|_{0} +\delta h \right),
 \label{forceAA}
 \end{equation}
\begin{equation}
 F_B=(-m_p)\left( \frac{dB_z}{dz}\biggm|_{0} +\delta h \right)+\,2 m_{in {\rm B}} \left( \frac{dB_z}{dz}\biggm|_{0} +\delta h \right), 
 \label{forceBB}
 \end{equation}
When the quantities represented in equations\,\eref{forceAA} and \eref{forceBB} are added and averaged, we obtain
\begin{equation}
\left|\frac{F_A +F_B}{2 }\right|=m_p \frac{dB_z}{dz}\biggm|_{0} - (m_{in {\rm A}}+ m_{in {\rm B}})\delta h - \frac{dB_z}{dz}\biggm|_{0}(m_{in {\rm A}}- m_{in {\rm B}}).
\label{avg-force}
\end{equation}
The first term in equation\,\eref{avg-force} is larger than the second and third because the permanent moment is more than two orders larger than the induced, and $\delta h \sim \frac{dB_z}{dz}\biggm|_{0}$. 
The acceleration of the TM due to magnetic force, neglecting the smaller terms in equation\,\eref{avg-force}, is
\begin{equation}
|\Delta a|=\frac{m_p}{M}\frac{dB_z}{dz}\biggm|_{0} ,
\label{inv-acc}
\end{equation}
where $M\sim 1$\,kg is the mass of the TM.

\section{Shielding \label{shield}}
Magnetic shielding is the use of a high permeability material like Mu-metal (see section\,\ref{mumetal}) to enclose a volume of space and reduce magnetic field inside the enclosed space by concentrating the flux within the material. Shielding factor is the ratio of the (external) magnetic field without the shield to the field at the center of the shield.
Magnetic shielding is more challenging than electrostatic shielding because of the lack of magnetic monopoles that are mobile in the shield. It is like electrostatic shielding using a dielectric. Historical accounts of the developments in magnetic shielding, including the  derivations of transverse and axial shielding factors of multiple cylinders, are presented in \cite{sumner,mager1,mager2}.

For a thin spherical shell in  a uniform external magnetic field, the field inside is constant with its direction aligned with that of the external field\;\cite{maxwell,stratton}. 
\begin{equation}
\fl S_{\rm sphere}=1+\left(\frac{2}{3}\right)\frac{(\mu-1)^2}{ \mu}\left(\frac{t}{r}\right)-\frac{(\mu-1)^2}{ \mu}\left(\frac{t}{r}\right)^2+\mathcal{O}\left(\frac{t}{r}\right)^3
\approx 1+\frac{ 2 t}{3r}\mu,
\label{sphere}
\end{equation}
 where $t$ is the thickness of the shield, $r$ is the radius of the shell. For large values of relative permeability ($\mu \gg 1$), the approximation for equation(\ref{sphere}) is valid. For Mu-metal, we assume an incremental relative permeability of $\mu=50,000$. For an infinitely long cylindrical shell with its axis transverse to the external field the shielding factor is:
 \begin{equation}
 S_{\rm cyl}=\frac{1}{4\mu}\left((1+\mu)^2-\frac{r_{\rm in}^2}{r_{\rm out}^2}(1-\mu)^2\right),
 \label{cyl}
 \end{equation}
where $r_{\rm in}$ and $r_{\rm out}$ are the inner- and outer-radius of a cylindrical shell.
Equations (\ref{sphere}) and (\ref{cyl}) are useful for validating the finite element analysis (FEA) presented in section\,\ref{fea}. Equation (\ref{cyl}) will serve to establish the ratio of length to radius of a cylinder that fits the definition of an infinitely long cylinder which is useful for estimating errors in numerical simulations.

In the following section, we use approximate analytic formulae\,\cite{sumner, mager1, mager2} to obtain the shielding factors for both transverse and axial fields. We apply these formulae to multiple shields to help us with first order estimates for shielding factors as a starting point for a thorough FEA study, see table\,\ref{tab2}.
A rough estimate of the gradient is obtained by dividing the field at the center of the inner shield by the radius of the shield. In section\,\ref{fea} we  use FEA to obtain relatively accurate estimates of the gradient.

\subsection{Shielding material: Mu-metal \label{mumetal}}
Mu-metal is a ferromagnetic alloy (approximately $ 75\% \rm{Ni}, 15\% \rm{Fe}, 5\% \rm{Cu}, 2\% \rm{Cr}$, etc.), that is widely used in a variety of  shielding applications. Here, we'll assume an incremental relative permeability of $\sim 50,000$ in an external field of $5\times 10^{-5}$~T for shields with thickness of 2 and 3~mm.
Another material available commercially is Metglas\,\cite{metglas}, an amorphous alloy. While its  permeability can be as high as $10^6$, it is available only in strips $\sim 15$ to $20\,\mu{\rm m}$ thick. The problem of low net permeability owing to the gaps between strips has been solved by using multiple sets of strips in a single shield layer, some running axially and some circumferentially. A single-layer shield with 6 laminations of Metglas was used to create a shield of radius 0.61 m and only $122\,\mu m$ thick with axial and transverse shielding factors (static and without degaussing) of 267 and 1500\,\cite{malkowski}. This sort of performance would require a shield of mu-metal to be several mm thick. 

Although multiple layers would have to be carefully held in place, e.g., use adhesive, to avoid the slightest unpredictable gravitational signal, Metglas is a shielding option well worth investigating in future. A third shielding material is Nanovate\,\cite{nanovate}, a nanocrystalline ferromagnetic coating whose magnetic properties are insensitive to physical deformation and which has a permeability comparable to that of Mu-metal for fields two orders of magnitude higher than the Earth's field. However, in Earth's magnetic field, its permeability is a factor of 5 smaller than that of Mu-metal.
As the external field becomes large $\sim 1$\,T  Mu-metal becomes saturated and the relative permeability approaches $\sim 1$, yielding a very low shielding factor. In such cases, Nanovate is a better alternative. In this paper, we'll use Mu-metal for all shield configurations. 

The permeability of a Mu-metal shield depends on its thermal and stress history. The values of permeability of the shield quoted by manufacturers are often obtained from small samples just after they have been annealed. The permeability is reduced by plastic strain and these reductions can be restored only by re-annealing\;\cite{sumner}.
During launch, our shield will experience a high level of vibration. A solid empirically-based estimate of the on-orbit permeability, or testing, will be needed.

The relative permeability can be maintained at optimal values by  ``demagnetization," (also known as degaussing) at regular intervals because SR-POEM payload undergoes a series of inversions. 
This can be done magnetically, mechanically, and thermally. For SR-POEM we are considering the magnetic process, which requires current in a surrounding coil driven by an AC signal with a soft starting envelope and an exponential decay. It may be possible to demagnetize the outer shield after each inversion, although the thermal perturbation, even to the outer shield, must be considered. If the external field remains constant, then by demagnetizing one may attain a lower shielded field for a fixed shield mass, corresponding to a permeability as high as 350,000\,\cite{sumner}. As the SR-POEM flight proceeds, the external magnetic field changes both in magnitude and direction resulting in the shield having a lower effective permeability\,\cite{sumner-p}.

\subsection{Cylindrical shields: Analytic shielding factors \label{analytic} }

\subsubsection{Transverse applied field}

Sumner \etal\cite{sumner} derive a formula for the shielding factor for an infinitely long hollow cylinder in a uniform external transverse field $B_0$ by finding the magnetic scalar potential to solve Laplace's equation with appropriate boundary conditions. For a shield of thickness $t$ and radius $R$, assuming  $t\ll R$ and $\mu\gg 1$, the shielding factor is $S^T\approx \mu t/2 R $\,\cite{sumner}. This result is approximately true for a closed-end finite-length shield as well because the shunting of flux by the end caps approximately compensates for the extra flux at the ends. 

Sumner \etal generalizes this result for $n$ infinite concentric cylinders in a transverse uniform field with each shield satisfying the above assumptions for thickness and permeability, to obtain a shielding factor 

\begin{eqnarray}
\fl S^T_n=1+\sum_{i=1}^{n}S^T_i+\sum_{i=1}^{n-1}\sum_{j>i}^{n}S^T_i S^T_j\left(1-\left(\frac{R_i}{R_j}\right)^2\right)\nonumber \\
+\sum_{i=1}^{n-2}\sum_{j>i}^{n-1}\sum_{k>j}^{n} S^T_i S^T_j S^T_k \left(1-\left(\frac{R_i}{R_j}\right)^2\right) \left(1-\left(\frac{R_j}{R_k}\right)^2\right)\nonumber \\
+ \cdots +
S_1^T \cdots S_n^T\left(1-\left(\frac{R_1}{R_2}\right)^2\right) \cdots 
\left(1-\left(\frac{R_{n-1}}{R_n}\right)^2\right),
\label{strans}
\end{eqnarray}
where $R_i$ is the  radius of the $i^{\rm th}$ shield, $n$ is the total number of shields, and $S^T_i\equiv \mu_i t_i/2 R_i $ is the transverse shielding factor corresponding to the  $i^{\rm th}$  shield with thickness $t_i$ and relative permeability $\mu_i$. 
In equation\,\eref{strans}, the term with $p$ sums includes a total of  
${{n}\choose{p}}$   terms.

\subsubsection{Axial applied field}

For a single, infinitely long, shield in a uniform axial external field, the shielding factor is $\sim 1$, i.e., there is no shielding. A closed-ended cylinder can be approximated as an ellipsoid of matching major and minor axes. When the length to radius ratio $a = L/R \gg 1$ and the transverse shielding factor $S^T \gg 1$, the shielding factor in an axial field is \cite{mager1}
\begin{equation}
S_{A} \simeq 1 - \frac{1}{a} + \frac{1}{a^2} + 16\,\frac{S^T (\log (a)-1)}{a^2} \rightarrow 1 \qquad \rm{as} \quad a\rightarrow\infty\, .
\end{equation}

For the axial shielding factor, a more empirical approach is needed. Sumner \etal assume a distribution of ``magnetic charge" with free parameters $\alpha$ and $\beta$, and adjust $\alpha$ and $\beta$ for the best fit to experimental shielding factors. They state that their results are in close agreement with Mager's estimates derived using ellipsoids \cite{mager1}. The axial shielding factor for a closed single cylinder of finite length $L_i$ is \cite{sumner}

\begin{equation}
S^A_i=1 + \frac{\mu t_i}{2 R_i }\left(\frac{ 2 K_i}{1 + a_i +\alpha a_i^2/3}\right),
\label{singleshieldaxial}
\end{equation}
where $a_i=L_i/R_i$ is the aspect ratio and $K_i$ represents the functional form of the scalar potential due to the ``magnetic charge" on the end caps up to a scaling factor,

\begin{equation}
\fl K_i=\left(1+\frac{1}{4 a_i^3}\right)\beta -\frac{1}{a_i}+2\alpha\left[\log\left(a_i+\sqrt{1+a_i^2}\right)-2\left(\sqrt{1+a_i^{-2}}-a_i^{-1}\right)\right].
\label{magcharge}
\end{equation}
Empirical values of $\alpha\sim 1$ and $\beta\sim 2$ are in good agreement with measured values of axial shielding \cite{sumner}.

The axial shielding factor for a set of $n$ nested cylinders may be obtained from equation\,\eref{strans} by making the following changes:
\begin{equation}
\left(\frac{R_i}{R_j}\right)^2 \rightarrow \frac{L_i}{L_j} \quad {\rm and} \quad S^T_i\rightarrow  S^A_i,
\label{saxial}
\end{equation}
where equation\,\eref{singleshieldaxial} represents the expression of axial shielding factor $S^A_i$ due to a single shield.

\begin{figure}[\h]
\begin{center}
\includegraphics[width=.45\textwidth]
{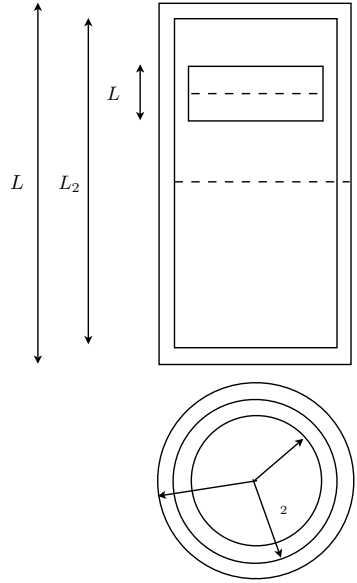}
\caption{Mu-metal shield comprising multiple(two or three) concentric cylinders. The closed innermost shield is off-set at a distance $z$ from the center of the outer shield. The outer shields may be closed or open. Two adjacent concentric shields have an separation $d=R_j-R_i$. The radii and lengths of the shields used are given in table\,\ref{tab2}.}
\label{fig2}
\end{center}
\end{figure}

Equations\,\eref{strans} and \eref{saxial} for the shielding factors, which are derived using recursion relations, provide information about the fields only at the centers of the cylinders. These formulae  are useful for comparing the shielding factors obtained using FEA. Because for SR-POEM, in addition to the field at the center, we also need to know the field gradient and its variation within a substantial volume inside the innermost cylinder as shown in figure\,\ref{fig2}.

\begin{figure}[\htb]
\begin{center}
\includegraphics[width=.85\textwidth]{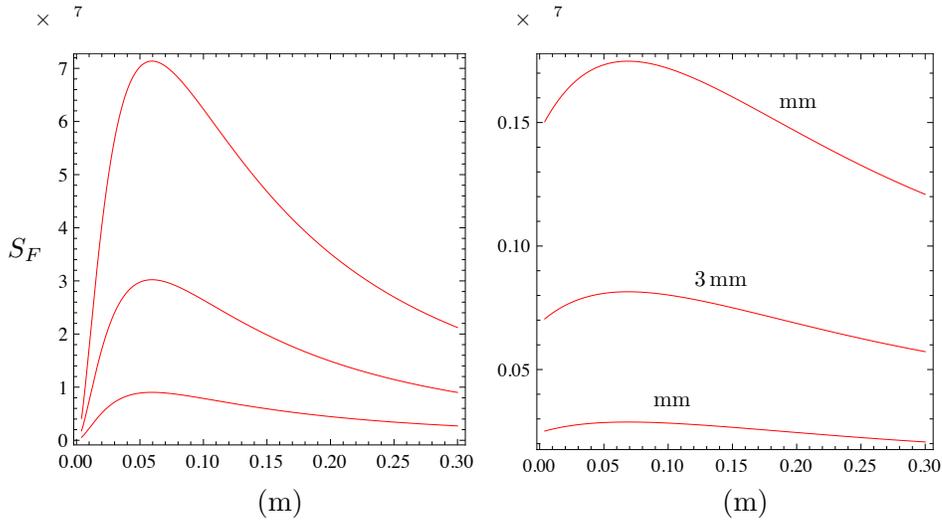}
\caption{Shielding factor as a function of radial separation between the shields --- analytic approximation: Three-shield configurations for shield thickness  2, 3 and 4 mm -- more thickness yields higher shielding efficiency. The radius of the inner shield is held constant at 10.5\,cm. $a$) For transverse field. $b$) For axial field. Shield lengths, inner to outer: 10, 98, 100 cm. Two-shield configuration yields similar results but the shielding factor is less by a factor of 60. }
\label{fig3}
\end{center}
\end{figure}

\begin{figure}[\h]
\begin{center}
\includegraphics[width= \textwidth]{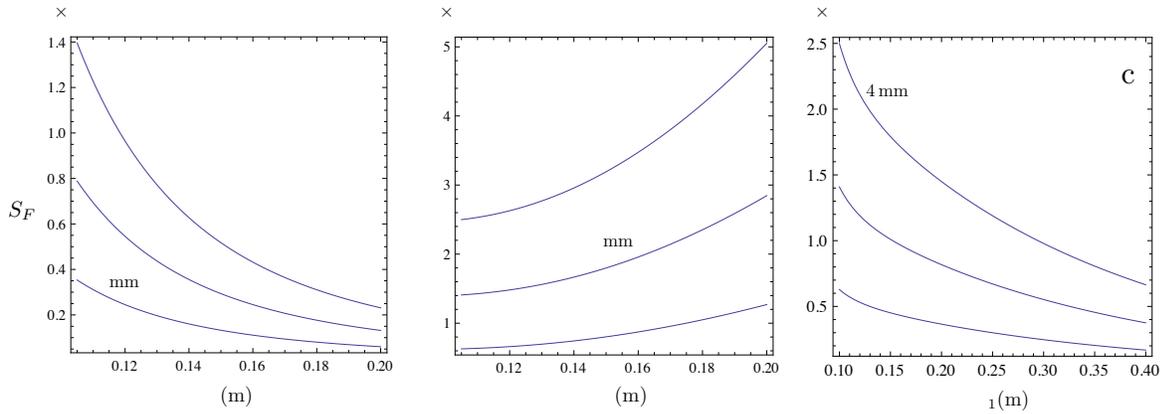}
\caption{Analytic approximation to shielding factor as a function of shield radius and length: For two-shield configurations with shield thickness  2, 3 and 4 mm. The separation between two successive shields is held constant at 1 cm. Permeability is 50,000. $a$) For transverse field, as a function of radius  $b$) For axial field, as a function of radius. $c$) For axial field, as a function of length of the innermost cylinder.}
\label{fig4}
\end{center}
\end{figure}

The largest dimension of the SR-POEM test mass assembly is less than 20\,cm. The maximum diameter allowed for the experimental package depends on the choice of the sounding rocket. To obtain analytic estimates using equations\,\eref{strans} and \eref{saxial}, we fix the diameters of the outermost shield to be 35.6\;cm and the innermost shield to be 21\;cm, see figure\,\ref{fig2} and table \ref{tab2}. For the case of three shields, the diameter of the middle shield is set to 25\;cm. The shielding factor as a function of radial separation and length (axial case only) of the shields are plotted in figures\,\ref{fig3} and \ref{fig4}  respectively.

Figure\,\ref{fig3} shows the variation of shielding factor as a function of shield separation for a three-shield configuration, which  includes the test case of shields 1, 2 and 3 in table\,\ref{tab2}. The radius of the innermost shield is held constant. 
For all thicknesses, the axial shielding factor is less by a factor of 10--40 than the transverse shielding factor for ordinary shield separations.
The transverse shielding factor for 3\,mm thickness is $\sim 6.0\times 10^{6}$ for a shield separation of 1\,cm. 
Also, the variation in transverse shielding factor with separation is steeper than the variation of axial shielding factor for separations of $\sim 1$ -- 5\,cm. The maximum shielding factors (both axial and transverse) for all thicknesses correspond to a shield separation of $\sim 5$\,cm. 
Axial shielding has not benefited from the middle shield in our three-shield configuration, because $L_2\approx L_3$, where $L_2$ and $L_3$ are the lengths of the middle and outer shields, see equation\,\eref{saxial}.

When the middle shield is removed, we find that the axial and transverse shielding factors are comparable  $\sim 1.4\times 10^{6}$ for our chosen configuration ($L_3/L_1\sim 5.5 $). If the radii of the shields are varied such that the distance separating them is held fixed, then the transverse shielding factor decreases --- proportional to the inverse square and cube of radii (to first order) for two- and three-shield configurations. For the axial case (see equation\,\eref{singleshieldaxial}), the shielding factor depends on the magnetic charge which is roughly a function of the cube of the aspect ratio (see equation\,\eref{magcharge}) and so the shielding factor increases with radius, see figure\,\ref{fig4}.

A comparative analysis of the analytic approximation for the gradient inside the inner shield and the corresponding FEA (section\,\ref{fea}) for 3\,mm thick shields (see table\,\ref{tab2} for shield geometry) is presented in table\,\ref{tab3}. In the following section we will use FEA to model the finite cylinders in uniform transverse and axial magnetic fields.

\begin{table}
\caption{ \label{tab2} Dimensions (cm) of Mu-metal shields with varying thicknesses of  2--4\,mm }
\begin{indented}
\item[]
\begin{tabular}{llll}
\br
Shield (i)  \hspace{4cm} & 1 \quad & 2\quad & 3\\
\mr
Radius ($R_i$)\hspace{2.0in} & 10.5 & 12.5  & 17.8 \\
Length  ($L_i$) & 18  & 100 & 150\\
Center & $(0,0,30)$  &  $ (0,0,0) $ & $ (0,0,0) $\\
\br
\end{tabular} 
\end{indented}
\end{table}


\Table{\label{tab3} Comparison of magnetic field and gradient: Analytic and FEA estimates for the shield thickness of 3\,mm.}
\br
&&\centre{1}{(cm)}&\centre{2}{Analytic$^{\rm 1}$}&\centre{2}{FEA$^{\rm 2}$}\\

\ns
&&\crule{1}&\crule{2}&\crule{2}\\
Shields & $B_0$ & $R_j -R_i$ $^{\rm 3}$ & $B_{in}$  & $dB/dz$ $^{\rm 4}$ & $B_{in}$  & $dB/dz$ $^{\rm 5}$\\

\mr

3 & Ax $^{\rm 6}$       & 2 \& 5.3 & $17.5\times 10^{-12}$ & $9.75\times 10^{-11}$ & $50.0\times 10^{-12}$ & $ 8.00\times 10^{-11}$ \\
3 & Tr                  & 2 \& 5.3 & $4.07\times 10^{-12}$ & $3.87\times 10^{-11}$ & $1.00\times 10^{-12}$ & $1.00\times 10^{-11}$  \\
2 & Ax                  & 7.3      & $5.33\times 10^{-10}$  & $2.96\times 10^{-9}$  & $10.0\times 10^{-10}$  & $50.0\times 10^{-10}$   \\
2 & Tr                  & 7.3      & $3.21\times 10^{-10}$ & $7.44\times 10^{-9}$ & $1.80\times 10^{-10}$ & $5.00\times 10^{-10}$  \\
\br
\end{tabular}

\item[]$^{\rm 1}$ Applicable only for finite length shields with closed ends for axial field and  infinitely long shields for transverse field.
\item[]$^{\rm 2}$ Shield 3 has open ends. $B_{out}=5\times 10^{-5}$\,T , the Earth's magnetic field.
\item[]$^{\rm 3}$ See table\,\ref{tab2} for the dimensions of the shield configuration.
\item[]$^{\rm 4}$ $B_{in}/r_{in}$ for transverse and $B_{in}/L_{in}$ for axial case. See table\,\ref{tab2} for the dimensions of the shield configuration.
\item[]$^{\rm 5}$ Maximum value of the gradient inside the inner shield.
\item[]$^{\rm 6}$ Ax, Tr : axial and transverse directions.
\end{indented}
\end{table}

\section{ Finite element analysis (FEA) of two- and three-cylinder shields \label{fea}}
We subjected various combination of shields summarized in table \ref{tab2} to FEA in which we simulated the application of uniform external transverse and axial magnetic fields.  We found the field and field gradient using the procedure below. 
\begin{enumerate}
\item {A FEA in COMSOL\,\cite{comsol}, with the finest mesh refinement available, and settings optimized for highest numerical accuracy.}
\item{The field inside the shield exhibits numerical noise associated with the spatial discretization. A fifth order polynomial is fit to the magnetic field along a line inside the shield; the analytic derivative gives the gradient.}
\item{An overall volume is set such that its linear dimensions are $\sim 50$ times the shield geometry along each axis.}
\item{A magnetic field $B_0=5\times 10^{-5} $ T (representing the Earth's field), is applied in turn along the axial and transverse directions.}
\item{The Mu-metal shields have an incremental permeability $\mu_r=5\times 10^4$.}
\item{The shield density, used only in calculating the mass, is taken to be that of Mu-metal, $\rho_\mu$=8750\,kg/m$^3$.}
\end{enumerate}

First, we validate the finite element analysis (FEA) using known exact solutions. For a spherical shell of thickness $3$ mm and radius $20$ cm, equation\,\eref{sphere} yields a shielding factor $500.999$. With COMSOL, we get $500.932$. 
For a 3mm thick infinite long cylinder in a transverse magnetic field, using equation(\ref{cyl}), we obtain a shielding factor 370.423 and with COMSOL we get 372.884. We approximated an infinitely long cylinder by setting length and radius of the cylinder to 3~m and 0.2~m respectively with $l/r\sim 15$. The estimated numerical error is $\sim 0.5\% $. For all our simulations the solutions converged (less than $10^{-3}$) under 30 iterations.

For a cylindrical shield, the attenuated field along the axis has the direction of the external field if the shield is parallel to or perpendicular to the external field. Away from the axis, this is not the case. 
Therefore, we consider all three magnetic field components at every point inside the shielded volume.  The notation for the plots that describe the gradients is as follows: When the shield is placed in a uniform external field, say $\vec{B}\equiv B_0 \hat{k}$, the field
has components $(B_x,B_y,B_z)$ at every point inside, although most of the contribution comes from the $z$\;component. 
We apply an external axial or transverse field separately for each of the shield configurations under investigation.
A representative set of plots corresponding to a case study is shown in figure\,\ref{fig6}.
The top row of plots represents the variation of the $x$, $y$ and  $z $ components of the field inside the shield and along the $x$ direction with $y=z=0$. The second row represents the component variations along the $y$ direction with $x=z=0 $ and the third row represents the variations along the $z$ direction with $x=y=0$. 
We considered the following cases for the FEA:
\begin{enumerate}
\item {Two- and three-shield configurations with closed ends}
\item {Two- and three-shield configurations with open-ended outer shield}

\end{enumerate}


We find consistently that for the three-shield configuration FEA, the transverse shielding factors are a factor of $\sim 150-180$ higher than the corresponding two-shield configuration. The analytic formula predicts a factor of $\sim 120-150$ higher. 
For both two- and three-shield configurations, the ratio of the field inside based on FEA and the analytic estimates is $B_{\rm FEA}/B_{\rm Analytic}\sim 3$. This  was true with both open- and closed-ended shields.

Similarly,  for the three-shield configuration FEA, we find that the axial shielding factors are a factor of $\sim 50-100$ higher than the corresponding two-shield configuration. The approximate analytic formula predicts a factor of $\sim 20-100$ higher. 
For both two- and three-shield configurations, the ratio of the field inside based on FEA to that obtained with the analytic approximation is $B_{\rm FEA}/B_{\rm Analytic}\sim 2$ for closed-ended shields.
We note that since most of our shield configurations have a length to radius  ratio, $l/r >2$, the results of Sumner \etal (1987) are not applicable\cite{sumner}.

\section{Results and discussion \label{results}}

We have investigated two- and three-shield configurations with open- and closed-ended outermost shields. 
Analytic estimates and the FEA results for two- and three-shield configurations in a transverse field are comparable.
FEA results of all the shield configurations we considered, see table\,\ref{tab4}, consistently give a lower shielding factor for axial fields than for transverse fields for the same shield configuration. FEA results for axial shielding factor was also lower than the analytic approximations (transverse and axial) for all the shield configurations considered here.
Therefore, we concentrate on meeting the required axial shielding factor. 
We note that the geometry of most of the shields that we have considered remain off bounds of validity for analytic expressions.
The results of our simulations are presented in table\,\ref{tab4}. The plots of the field and gradient inside the innermost closed shield for a representative configuration in an axial external field (case~3) are given in figure\,\ref{fig6}. From similar plots corresponding to other cases, we obtain the values of maximum gradients; also given in the gradient column of table~\ref{tab4}.

Spacing between the shields can be increased to improve both the transverse and axial shielding factors, although the improvement in axial shielding factor is modest when compared to transverse shielding, see figure\,\ref{fig3}. Axial shielding factor can be enhanced by decreasing the length of the inner most shield. On the downside, the assumed thickness of 
3\,mm seems to be at least a factor of two larger than the commercially available sheets of Mu-metal. Our solution to this problem was to decrease the shield thickness and to move the inner shield close to the center of the outer shield.

\begin{figure}[\htb]
\begin{center}
\includegraphics[width= \textwidth]{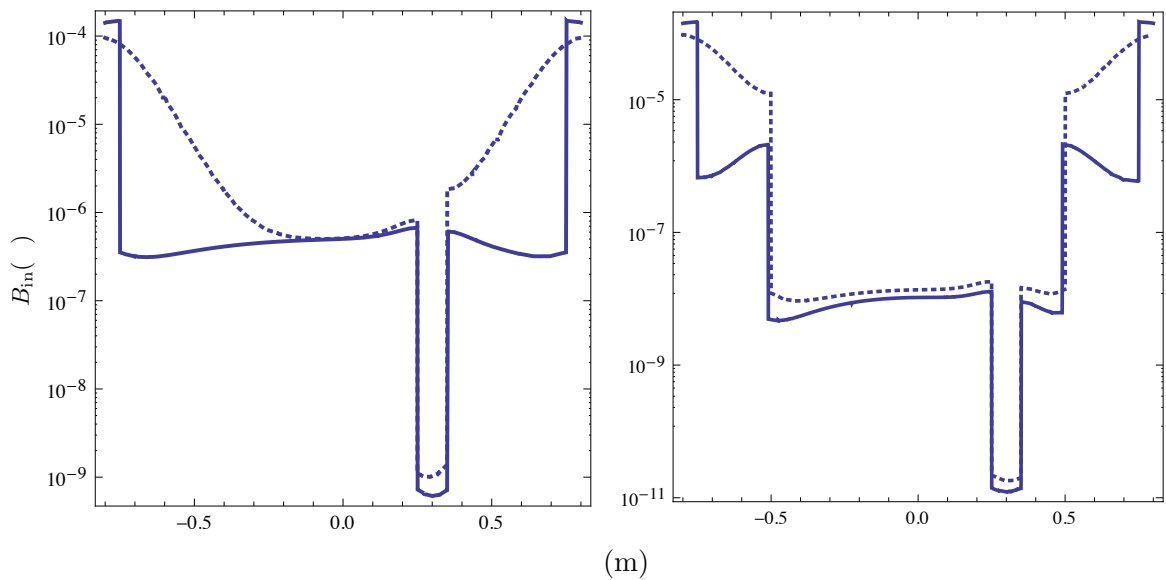}
\caption{ FE analysis: Internal field resulting from an axial applied field with the outer shield closed (solid line) and open (dashed).
a) For the two-shield configuration, the maximum axial shielding factor (along the direction of the external field, $z$ direction) is $\sim 5\times 10^4$. The minimum field at the center of the innermost cylinder is $\sim 10^{-9}$\;T for an external field of $5\times 10^{-5}$\;T directed along the z (axial) direction. b) For the three shields, the axial shielding factor is $\sim 5\times 10^6$ and the minimum field at the center of the innermost cylinder is $\sim 10^{-11}$\;T. All the shields considered are 3 mm thick. The open ended outermost shield does not affect the transverse shielding factors.}
\label{fig5}
\end{center}
\end{figure}

\begin{figure}[\htb]
\begin{center}
\includegraphics[width= 0.95\textwidth]{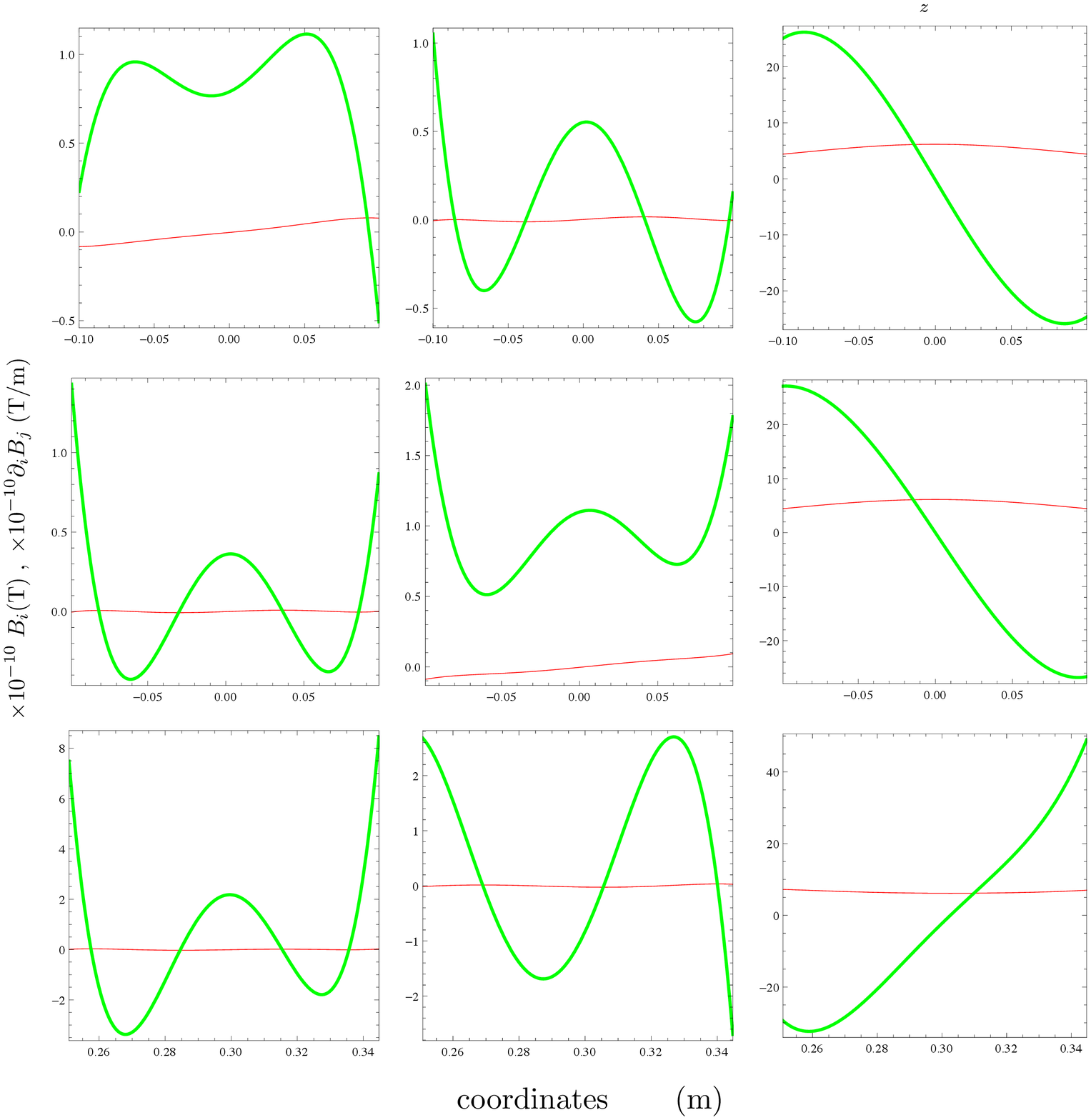}
\caption{Results of FE analysis: Case 3, table\,\ref{tab4}). The concentric shields 1 (inner -- thickness 3\;mm) and 3 (outer -- thickness 2\;mm) are placed in an external field $5\times 10^{-5}$\;T directed along the z\;(axial) direction.
The maximum value of the field at the center is $\sim 7.0 \times 10^{-10}$\;T.  The value of the maximum gradient within the inner cylinder is $\sim 2.0 \times 10^{-9}$\;T/m.  In each panel, the thick  and varying (sometimes wildly) curve corresponds to the gradient and, the thin and near constant curve corresponds to the field. For more on notations, please refer to section~\ref{fea}}
\label{fig6}
\end{center}
\end{figure}

The question of keeping the ends open plays a vital role in shielding axial fields. For transverse fields
the open ends of the outermost cylinder have no effect on the shielding factor, both in terms of the maximum field inside and the gradient. This is as we expect, for  field and gradients evaluated sufficiently far away from the ends. For the axial case, the maximum field inside the innermost cylinder is slightly lower with the ends of the outer shield open (figure\;\ref{fig5}), and the maximum gradient is slightly larger than in the closed-ended case.

The values of the field gradients fare better (decrease) if one chooses optimal values for the length ratios with an open shield configuration compared to the closed outer shield configuration\;\cite{paperno,nagashima}.

\begin{table}
\caption{\label{tab4} Gradient in the several magnetic Shield and Coil Configurations analyzed with FEA}
\begin{indented}
\item[] 
\begin{tabular}{@{}lcllllll}
\br
$\#$  & Shield S1 & Shield S2  & Shield S3   & Offset$^{\rm 1}$    &$B_{0}$$^{\rm 2}$   & $dB_i/dz$$^{\rm 3}$    & Mass$^{\rm 5}$\\
   & (cm)        &  (cm)        & (cm)          & (cm)    &         & (T/m)        & (kg)\\
\mr
1  & 21\;18\;0.3$^{\rm 4}$ &25\;100\;0.3 &35.6\;150\;0.3 &  30         & Tr & $1.0 \times 10^{-11}$   & 76.0\\
   & closed      &   closed       & closed        &               & Ax      & $5.0 \times 10^{-11}$        \\
\mr
2  &            &25\;100\;0.3&35.6\;150\;0.3 &  30        & Tr & $1.0 \times 10^{-11}$   & 70.8\\
   &      "       & closed       & open          &              & Ax     & $8.0 \times 10^{-11}$                \\
\mr
3  &             & N/A         &35.6\;150\;0.3 &  30        & Tr & $5.0 \times 10^{-10}$   & 52.8\\
   &      "       &             & closed          &              & Ax      & $2.0 \times 10^{-9}$              \\
\mr
4  &             & N/A         &35.6\;150\;0.3 &  30        & Tr & $5.0 \times 10^{-10}$   & 47.6\\
   &      "      &             & open         &              & Ax      & $5.0 \times 10^{-9}$              \\
\mr
\end{tabular}
\item[] $^{\rm 1}$ Offset of geometric centers of S1 and S3 along the axial direction.
\item[] $^{\rm 2}$ Assumes an applied field of $B_0 = 5.0 \times 10^{-5}$ T along the transverse (Tr) or axial\,(Ax) directions.
\item[] $^{\rm 3}$ The table shows the largest component of $dB/dz$. This requirement is $\partial B_i/\partial z < 10^{-9}$\,T/m.
\item[] $^{\rm 4}$ Diameter, length and thickness of shield.
\end{indented}
\end{table}

\section{Conclusion}
In this paper, we have discussed ways to reduce the magnetic acceleration of the SR-POEM TMs. The source of this acceleration is the interaction of the remanent moment of the TMs with the gradient of the field inside the shielded volume. Some components of this acceleration do not cancel with inversion of the payload.
As a result, we set a requirement that the magnetic field inside the innermost shield for SR-POEM have value of $z$-derivative no more than $10^{-9}$\;T/m. 
In order to address this problem, we studied various cases involving multiple shields of different thickness and shields with open and closed ends. Most of the shield configurations that we investigated here, see table\;\ref{tab4}, will help reduce the magnetic field gradient to a value $\le 10^{-9}$~T/m and thereby meet the SR-POEM requirement of measuring $\eta$ to an accuracy of  $1\times10^{-17}$.

In table\;\ref{tab4}, cases 1 and 2---three-shield configurations---meet SR-POEM mission requirement (gradient inside the innermost shield is almost two orders of magnitude smaller than the limiting value). Since cases 3 and 4 are less heavier, they  are more preferred. 

FEA presented here alone is not adequate enough for selecting the shield configurations in cases 3 and 4. Previous efforts in reducing magnetic disturbance within a volume comparable to SR-POEM test-case  have largely come from the atomic clock community\cite{burt,heavner}. For example, NIST-F2 has a shield configuration very similar to cases 1 and 2. Heavner \etal report a measured magnetic field uncertainty of $\sim 50$~pT by measuring  $|3,1\rangle\rightarrow|4,1\rangle$ Ramsey 
fringes\cite{heavner}. They report a uniform field of $\sim 10^{-7}$~T. For SR-POEM, in addition to using multi-layer shields we may have to demagnetize the shields before each inversion. This is because the changing direction of Earth's magnetic field will reduce the permeability of Mu-metal by a factor of 10 or so.
Often, running a 60~Hz AC current of $\sim 100$ ampere-turns over few tens of seconds is adequate for degaussing a shield\cite{jefferts-p, heavner-p}. In space experiments deguassing may be implemented with a DDS circuit using DC current drawn from a battery.
By demagnetizing before each inversion we should be able to maintain the value of incremental permeability of Mu-metal to above $\mu=50,000$.

Commercially available atomic magnetometers can measure magnetic fields less than $0.1$~nT with a sampling rate of $\sim$kHz \cite{shah,shah-p}. These devices are tiny enough to be placed above and below SR-POEM TMs for taking measurements during the experiment phase. This is an option worth considering to make sure that magnetic field and gradients of the field are kept below the limiting values.

We have also considered the case of active suppression with a coil wound on the outside of the inner shield  to counter the Earth's magnetic field. This approach adds another layer of complexity, but when used in conjunction with an atomic magnetometer (magnetic servo-loop), will provide very effective field cancellation.

\ack 
This work was supported by NASA grant NNX08AO04G. We thank Robert Reasenberg and James Phillips for their help in getting this work started and for their guidance and suggestions as the manuscript evolved. We also thank Neil Ashby, Jens Gundlach, Tom Heaver,Mauro Hueller, Steven Jefferts,John Mester, Riley Newman,Vishal Shah and Tim Sumner for useful discussions. 

We would like to extend our thanks to two anonymous referees for their valuable suggestions that have helped improve this paper.  



\section*{References}
\begin{thebibliography}{10}


\bibitem{hueller} Hueller M, Armano M, Carbone L, Cavalleri A, Dolesi R, Hoyle C D, Vitale S and Weber W J 2005 
\newblock Measuring the LISA test mass magnetic properties with a torsion pendulum
\newblock \CQG {\bf 22} S521-26

\bibitem{hueller-p} Hueller M, Private communication 2010

\bibitem{gundlach-p} Gundlach J, Private communication 2010

\bibitem{igrf} International Geomagnetic Reference Field, extrapolated from 600 km to 1500 km, http://www.ngdc.noaa.gov/geomag/WMM/back.shtml, see Calculators, left-hand side.

\bibitem{mager1} Mager A J 1968 
\newblock Magnetic shielding efficiencies of cylindrical shells with axis parallel to the field
\newblock \JAP {\bf 39} 1914

\bibitem{mager2} Mager A J 1970 
\newblock Magnetic shields
\newblock {\it IEEE Trans. Magn.} {\bf MAG-6} 67-75 

\bibitem{malkowski} Malkowski S, Adhikari R, Hona B, Mattie C, Woods D, et al.
\newblock {\em Technique for high axial shielding factor performance of large-scale, thin, open-ended, cylindrical Metglas magnetic shields}
\newblock {\it Rev. Sci. Instrum.} {\bf 82} 075104 (2011).

\bibitem{maxwell} Maxwell J C 1904 
\newblock {\em Electricity and Magnetism}
\newblock  Oxford

\bibitem{mester} Mester J C and Lockhart J M 1996 
\newblock Remanent magnetization of instrument materials for low magnetic field applications
\newblock {\it Czech. J. Phys.} {\bf 46} (Suppl. 5) 2571-2

\bibitem{mester-p} Mester J C, Private communication 2010


\bibitem{nagashima} Nagashima K, Sasada I and Tashiro K 2003
\newblock Adaptive compensation of magnetic fields inside an open cylindrical magnetic shield
\newblock {\it \etal IEEE Trans. Magn.} {\bf 39} 3223-25 

\bibitem{paperno} Paperno E, Sasada I and Naka H 1999
\newblock Self-compensation of the residual field gradient in double-shell open-ended cylindrical axial magnetic shields
\newblock  {\it  IEEE Trans. Magn.} {\bf 35} 3943-45 


\bibitem{jdp}  Phillips~J D and Reasenberg~R D 2005
\newblock Tracking frequency laser distance gauge
 \RSI {\bf 76} 06450 (9 pages)
 
 \bibitem{rdr1}
 Reasenberg~R D, Lorenzini~E C, Patla~B R, Phillips~J D, Popescu~E M, Rocco R,
  and Thapa R  2011
\newblock A quick test of the wep enabled by a sounding rocket.
\newblock  \CQG {\bf 28} 094014(11pp)


\bibitem{rdr2} Reasenberg~R D and Phillips~J D 2010
\newblock A weak equivalence principle test on a suborbital rocket
\newblock  \CQG {\bf 27} 095005 (14pp)


\bibitem{rdr3}
Reasenberg~R D, Patla~B R, Phillips~J D, and Thapa R 2012
\newblock Design and characteristics of a wep test in a sounding-rocket
  payload.
\newblock \CQG  {\bf  29}  184013(18pp) 

\bibitem{rdr4}
Reasenberg~R D
\newblock  A new class of equivalence principle test masses, with application to SR-POEM
\newblock \CQG  {\bf  31}  175013(13pp) 

 
\bibitem{stratton} Stratton~J 1941 
\newblock {\em Electromagnetic Theory}
\newblock  McGraw-Hill


\bibitem{su} Su Y, Heckel B R, Adelberger E G, Gundlach J H, Harris M, Smith G L and Swanson H E   1994
\newblock New tests of the universality of free fall
 \newblock \PR D {\bf 50} 3614--36

\bibitem{sumner} Sumner~T J, Pendlebury~J M and Smith~K F 1987 
\newblock Conventional magnetic shielding
\newblock {\em J. Phys. D: Appl. Phys.} {\bf 20}, 1095-01 

\bibitem{sumner-p} Sumner~T J, Private communication 2010

\bibitem{thapa} Thapa R, Phillips J D and Reasenberg R D 2011
\newblock Subpicometer length measurement using semiconductor laser tracking frequency gauge
\newblock { \it Opt. Lett.} {\bf 36}  3759-61

\bibitem{metglas} { www.metglas.com}

\bibitem{nanovate} { www.integran.com}

\bibitem{comsol} { www.comsol.com}

\bibitem{burt} Burt~E A, Klipstein~W M and Jefferts~S R 2004
\newblock The cesium physics package design for the PARCS experiment
\newblock Frequency Control Symposium and Exposition, 2004. Proceedings of the 2004 IEEE International, pp71-79 

\bibitem{jefferts-p} Jefferts~S R, Private communication 2015

\bibitem{jefferts} Jefferts~S R, Shirley~J, Parker~ T E, Heavner~ T P, Meekhof~ D M,
Nelson~C, Levi~F, Costanzo~G, De Marchi~A, Drullinger~R, Hollberg~L, Lee~W D and 
Walls~ F L 2002
\newblock Accuracy evaluation of NIST-F1
\newblock Metrologia {\bf 39} 321-336



\bibitem{heavner}Heavner~ T P, Donley~E A,  Levi~F,
Costanzo~G, Parker~T E, Shirley~ J H, Ashby~N, Barlow~ and Jefferts~S R
\newblock First accuracy evaluation of NIST-F2
\newblock Metrologia {\bf 51} 174-182

\bibitem{heavner-p} Heavner~ T P, Private communication 2015

\bibitem{shah}Shah~V K and  Wakai~R T 2013
\newblock A Compact, High Performance Atomic Magnetometer for Biomedical Applications. \newblock Physics in medicine and biology, 58(22):8153-8161

\bibitem{shah-p} Shah~ V K, Private communication 2015

\end{thebibliography}
\end{document}